\documentclass[sigconf]{acmart}
\AtBeginDocument{%
  }

\usepackage{array}
\usepackage{float}
\usepackage{xcolor}
\usepackage{placeins}

\copyrightyear{2026}
\acmYear{2026}
\setcopyright{cc}
\setcctype{by}
\acmConference[UIST Adjunct '26]{The 39th Annual ACM Symposium on User Interface Software and Technology}{November 02--05, 2026}{Detroit, MI, USA}
\acmBooktitle{The 39th Annual ACM Symposium on User Interface Software and Technology (UIST Adjunct '26), November 02--05, 2026, Detroit, MI, USA}
\acmDOI{10.1145/3830397.3842454}
\acmISBN{979-8-4007-2855-6/2026/11}

\begin{document}

\title{Feeling \textsc{UISTful}: An Interactive Portrait of Scholarly Authorship, Readership, and the In-Between}


\author{Sophia W. Liu}
\affiliation{%
  \institution{University of California, Berkeley}
  \city{Berkeley}
  \state{California}
  \country{USA}
}
\email{sophiawliu@berkeley.edu}

\author{Shm Garanganao Almeda}
\affiliation{%
  \institution{University of California, Berkeley}
  \city{Berkeley}
  \state{California}
  \country{USA}
}
\email{shm.almeda@berkeley.edu}

\author{Max Kreminski}
\affiliation{%
  \institution{Cornell Tech}
  \city{New York}
  \state{New York}
  \country{USA}
}
\email{mkremins@cornell.edu}

\author{Bjoern Hartmann}
\affiliation{%
  \institution{University of California, Berkeley}
  \city{Berkeley}
  \state{California}
  \country{USA}
}
\email{bjoern@berkeley.edu}

\renewcommand{\shortauthors}{Liu et al.}

\begin{abstract}
We introduce \textsc{UISTful}, a system that turns reading activity into a collective portrait of a scholarly community. Readers explore a semantic globe of UIST papers and authors while the system records private reading traces that can be reviewed, reflected upon, curated, and published for others to replay. Inspired by the information flâneur, \textsc{UISTful} treats a reading trace as a camera through which readers frame and interpret what they read, casting reading as a creative and authorial process. Zooming out from individual traces reveals a bird's-eye portrait of the scholarly landscape, composed through the traces of many readers. Shared traces display the plurality of interpretations across the same scholarly corpus, inviting UIST to see itself as an interactive system of papers, authors, readers, and their exchanges.
\end{abstract}


\begin{teaserfigure}
 \includegraphics[width=\textwidth]{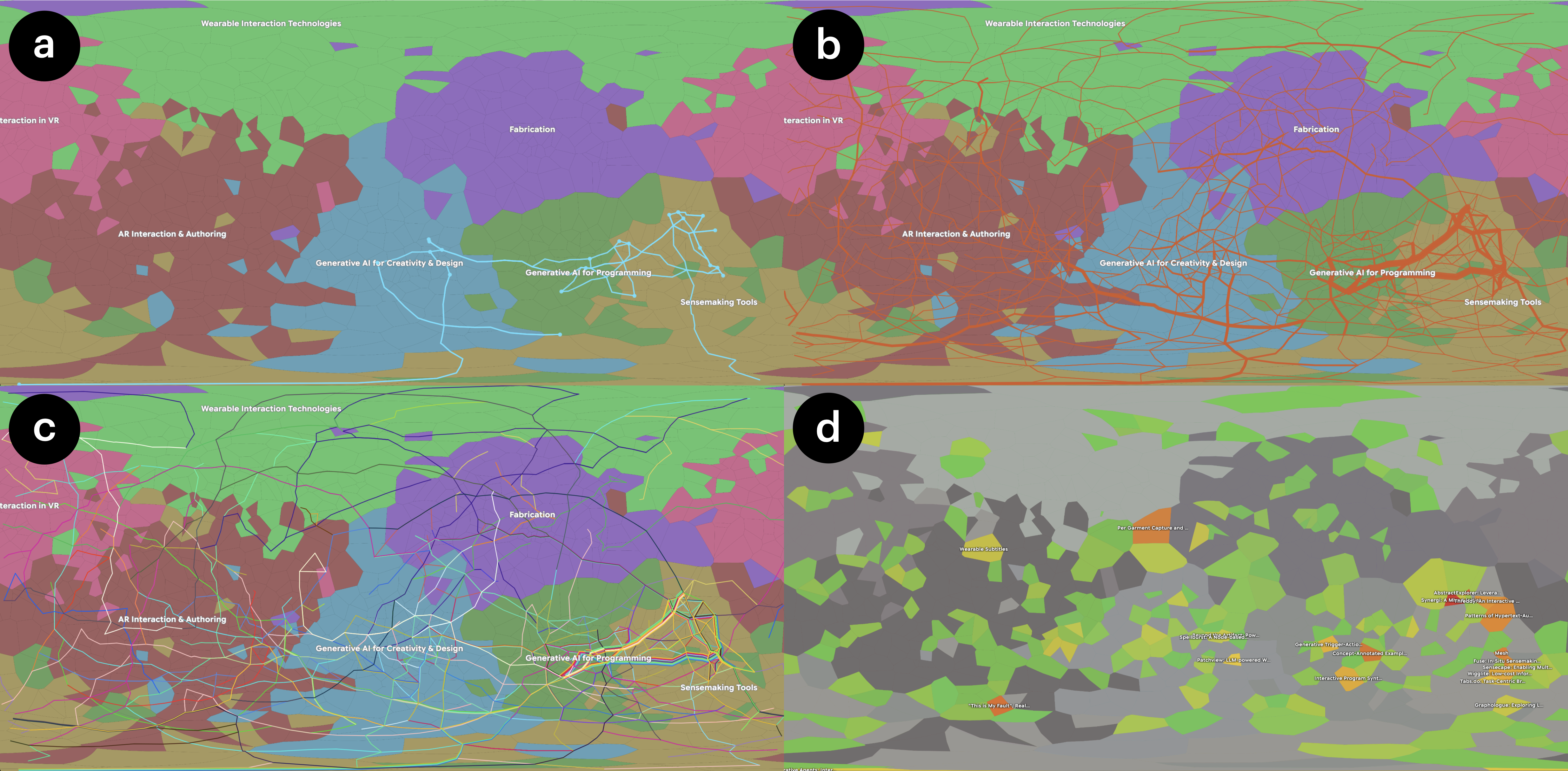}
    \caption{
The 2D equirectangular Map organizes UIST 2020--2025 full papers into seven semantic continents, with each cell representing a paper. Reading traces are shown in four views: (a) one reader's individual trace, (b) aggregate \textit{Paths}, with line thickness indicating traversal frequency, (c) individual \textit{Readers} in their distinct trace colors, and (d) \textit{Heat}, which colors each paper cell by visit frequency and labels the 20 most-visited papers. These views visualize reading activity from 54 unique readers over 31 days, spanning 247 paper nodes and 136 author nodes.}
    \Description{Four side-by-side views of the same 2D equirectangular map, divided into seven labeled semantic continents. Panel A shows one reader's trace as blue lines connecting visited paper and author labels. Panel B shows aggregate paths in orange, with thicker lines representing more frequent traversal. Panel C shows individual readers' paths in distinct colors, making overlaps and differences visible. Panel D colors paper cells by visit frequency and labels the 20 most-visited papers. The figure summarizes reading activity from 54 unique readers over 31 days, spanning 247 paper nodes and 136 author nodes.}
    \label{fig:maptraces}
\end{teaserfigure}

\maketitle

\section{Introduction}
Reading augmentation systems increasingly use LLMs and agentic workflows to perform reading and synthesis on the reader's behalf~\cite{creativeReading}. Many scholarly sensemaking systems instead scaffold readers in carrying out this work themselves across research texts at scale~\cite{SemanticReader, abstractExplorer}. Whether automating or assisting the process, these approaches provide powerful support for \textit{reading for transmission} through information extraction and synthesis, but offer comparatively less support for interpretive and \textit{creative reading}~\cite{creativeReading}.

One mode of such creative reading is the curiosity-driven exploration of wiki rabbitholing~\cite{goingDownTheRabbithole} and \textit{information flânerie}, which describe wandering through textual landscapes~\cite{informationCity} without a fixed destination, encountering information by bumping into it~\cite{informationFlaneur}. Like the camera of the urban flâneur~\cite{onPhotography}, a creative activity trace can preserve an otherwise ephemeral creative process for later revisitation~\cite{CATsDesignSpace}. Earlier work automatically captured browsing interaction histories to reduce effort in directed search~\cite{footprintsMaes}. More recent systems support the deliberate curation of research threads~\cite{Threddy}, ``paper gardens''~\cite{gardenOfPapers}, and shareable research trails~\cite{Semble} through the reader's explicit selection of sources and links. This suggests an opportunity to pair explicit curation with automatic, implicit capture so that the reader's path is recorded as a \textit{trace} they can revisit and layer interpretation onto. This approach is especially timely as LLMs and AI summaries compress reading into immediate answers and risk diminishing the curiosity, creativity, and critical eye of the flâneur~\cite{informationFlaneur}.

We introduce \textsc{UISTful}, a system for collectively capturing and composing an interactive portrait of a scholarly community of readers, authors, and those in between. Authors read to write, while readers exercise an authorial role by composing their own interpretations and syntheses across texts. At a conference, the readers are often also the authors of the papers circulating among them. This duality recalls the ``participant-observer dialectic'' of the urban flâneur, who is part of the very crowd they observe~\cite{painterOfModern}. In \textsc{UISTful}, readers navigate a semantic globe of recent UIST papers and authors, zoom into individual pages, and follow citation or authorship links. The system records each path as a private reading trace that readers can inspect, annotate, and curate before publishing. Published traces (Fig.~\ref{fig:publishedtrace}) appear in a public library, while anonymized full reader traces appear across the Globe, showcasing a plurality of readings and supporting reflection at both individual and collective levels. In effect, \textsc{UISTful} arms the information flâneur with a camera that encourages wandering and noticing while preserving each reading for later, even \textit{wistful}, viewing~\cite{onPhotography, informationFlaneur}. The result resembles a live wedding---or conference---portrait whose subjects are also its portraitists. Through this demonstration, we invite the UIST community, as researchers of interactive systems, to view itself as a living and observable whole~\cite{seeingWholeEarth}.

\section{\textsc{UISTful}: Interaction}
\textsc{UISTful} provides three main interfaces: the Globe for exploring papers and authors, \textit{My Trace} for reviewing and reflecting on one's own reading trace, and the \textit{Trace Library} for publishing and replaying community traces.

\subsection{Globe View: Reading}
The Globe serves as both an overview of the UIST community and an entry point for reading, situating individual papers and authors within the broader scholarly landscape.

\subsubsection{The \textsc{UISTful} Globe}
Inspired by spatial browsing interfaces like Artographer~\cite{Artographer}, the \textsc{UISTful} Globe (Fig.~\ref{fig:twotraces}) maps UIST 2020--2025 full papers and recurring authors onto seven semantic continents. Paper titles and abstracts are embedded and projected onto a sphere with spherical UMAP, authors are positioned at the centroid of their papers, and papers are grouped with \(k\)-means in embedding space into seven clusters named by an LLM. A Voronoi tessellation gives each paper a cell, while shortest paths over neighboring-paper Delaunay edges route citation and authorship links. The same layout is also available as a 2D equirectangular map. Readers rotate, pan, search, and zoom through the corpus much like Google Earth, with semantic zoom progressively revealing more labels~\cite{GoogleEarth, Sensecape}. By eliminating the edges of a planar map, the Globe presents the proceedings as a bounded yet continuously navigable whole, supporting an overview effect while preserving meaningful local neighborhoods~\cite{seeingWholeEarth}.

\subsubsection{Reading Interaction}
\begin{figure}
\centering\includegraphics[width=0.99\linewidth]{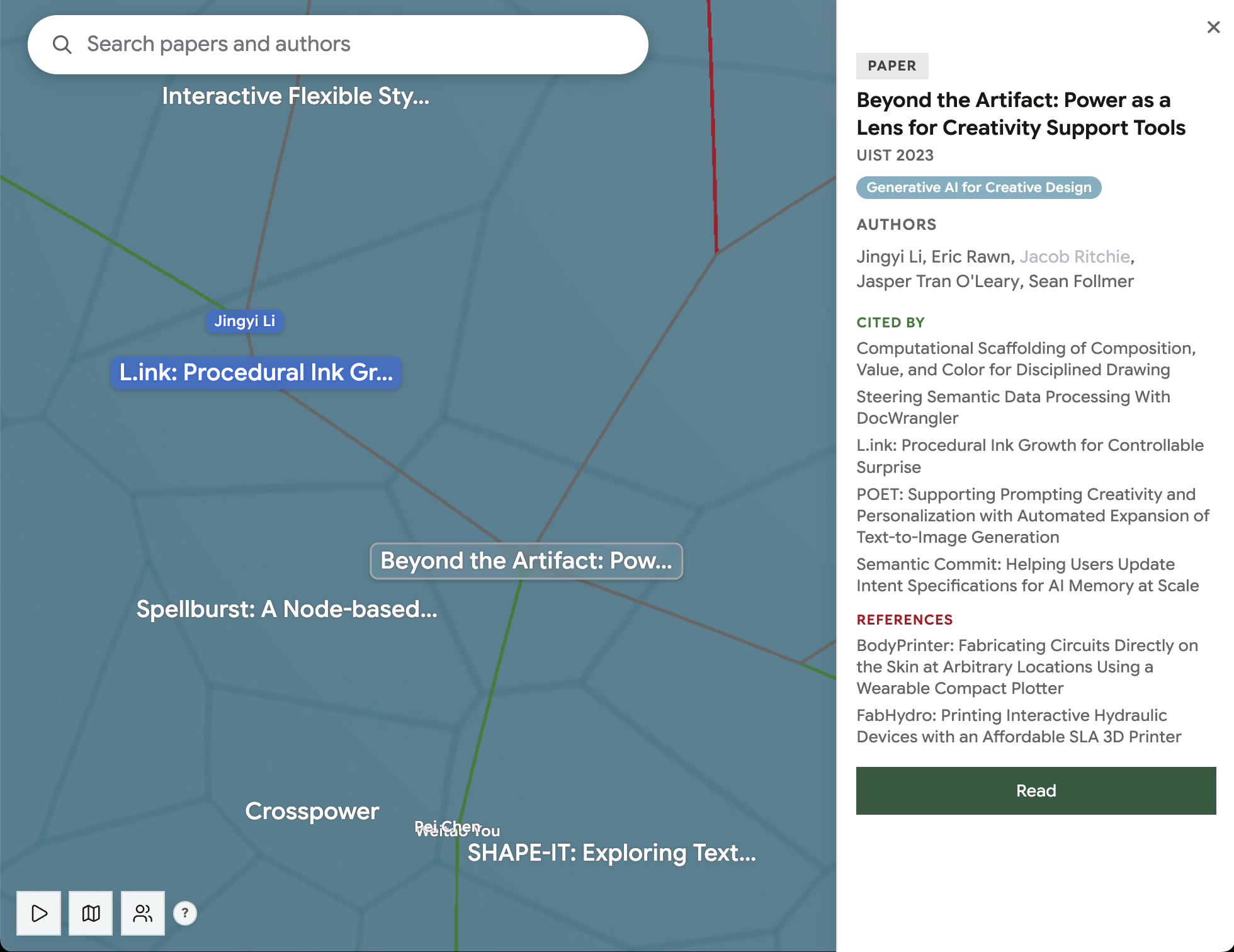}
    \caption{The \textsc{UISTful} Globe zoomed in on the selected paper, showing its links to neighboring papers and authors, with a side panel displaying metadata, authors, citations, and a button to read the full paper page.}
    \Description{A zoomed-in view of the UISTful Globe after selecting a paper label. The semantic map fills the left side, with nearby paper and author labels positioned across connected cells. The selected paper, ``Beyond the Artifact: Power as a Lens for Creativity Support Tools,'' is outlined on the Globe. A side panel on the right shows its title, year, semantic continent, authors, citing papers, and references, along with a Read button that opens the full paper page.}
    \label{fig:zoomedinpinned}
\end{figure}

\begin{figure}
\centering

    \setlength{\fboxsep}{0pt}
    \setlength{\fboxrule}{0.2pt}
    \fcolorbox{gray!20}{white}{
        \includegraphics[width=0.99\linewidth]{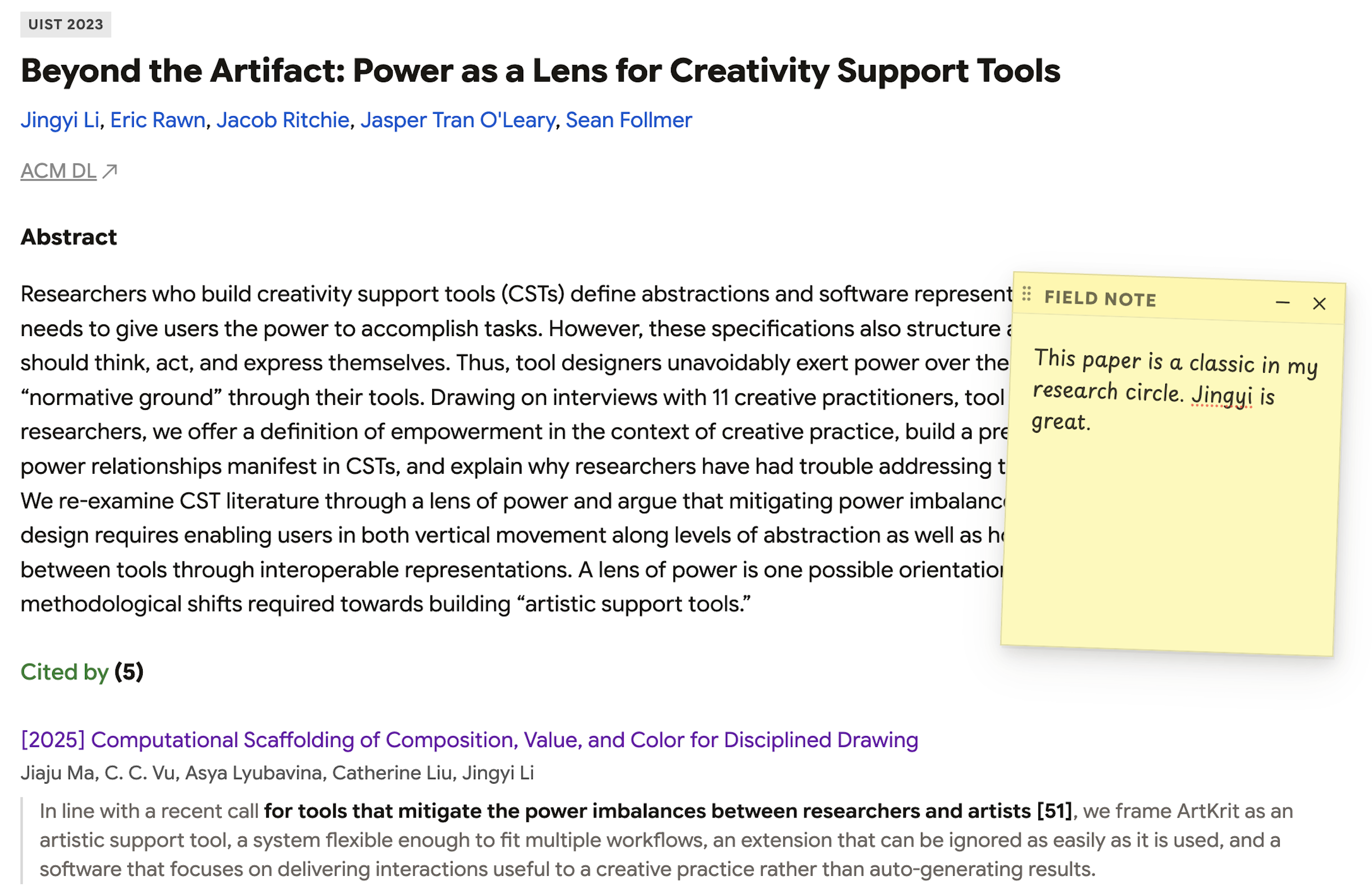}}
        
    \caption{A full paper page with linked authors and in-context citations, plus a \textit{Field Note} for reflection during reading.}
    \Description{A full paper page for ``Beyond the Artifact: Power as a Lens for Creativity Support Tools.'' The page shows the paper title, linked author names, an ACM Digital Library link, the abstract, and a cited-by section with in-context citation excerpts. A yellow sticky Field Note on the right contains a reader’s personal note written during reading.}
    \label{fig:readingview}
\end{figure}

Readers can wander across the Globe, search directly, or use \textit{I'm Feeling Lucky} to find a paper or author. Selecting a label opens a side panel with metadata and links to references, citing papers, and authors (Fig.~\ref{fig:zoomedinpinned}). Selecting \textit{Read} opens a full page with a yellow sticky \textit{Field Note} for in-reading reflection (Fig.~\ref{fig:readingview}). Paper pages include the title, author links, abstract, and in-context citations, while author pages list papers and collaborators. Readers follow blue hypertext links between pages while the system records each visit and traversed link in their private trace. Readers can also add external links to their trace when something outside \textsc{UISTful} comes to mind.

\subsection{Trace View: Sharing}
A \textit{reading trace} records the ordered paper and author pages a reader visits, including revisits and traversed links. Readers inspect and reflect on this private trace in \textit{My Trace}, then curate and publish selected traces to the \textit{Trace Library}.

\subsubsection{My Trace}
\textit{My Trace} uses the same coordinated Path, Tree, Graph, and notes interface shown in Fig.~\ref{fig:publishedtrace}, but in an editable form. Readers can scrub between steps in their trace, revisit in-reading \textit{Field Notes}, write retrospective, post-reading \textit{Reflection Notes}, bookmark steps, reveal latent (untraversed) citation and authorship edges, and add interpretive links between pages. They can then crop and title the trace, add an overall reflection, and either save it privately or publish it. Their current trace also appears on the Globe as visited paper cells and authors connected by the shortest paths between them (Fig.~\ref{fig:twotraces} and~\ref{fig:maptraces}-a).

\subsubsection{Trace Library}
\begin{figure}
\centering\includegraphics[width=0.99\linewidth]{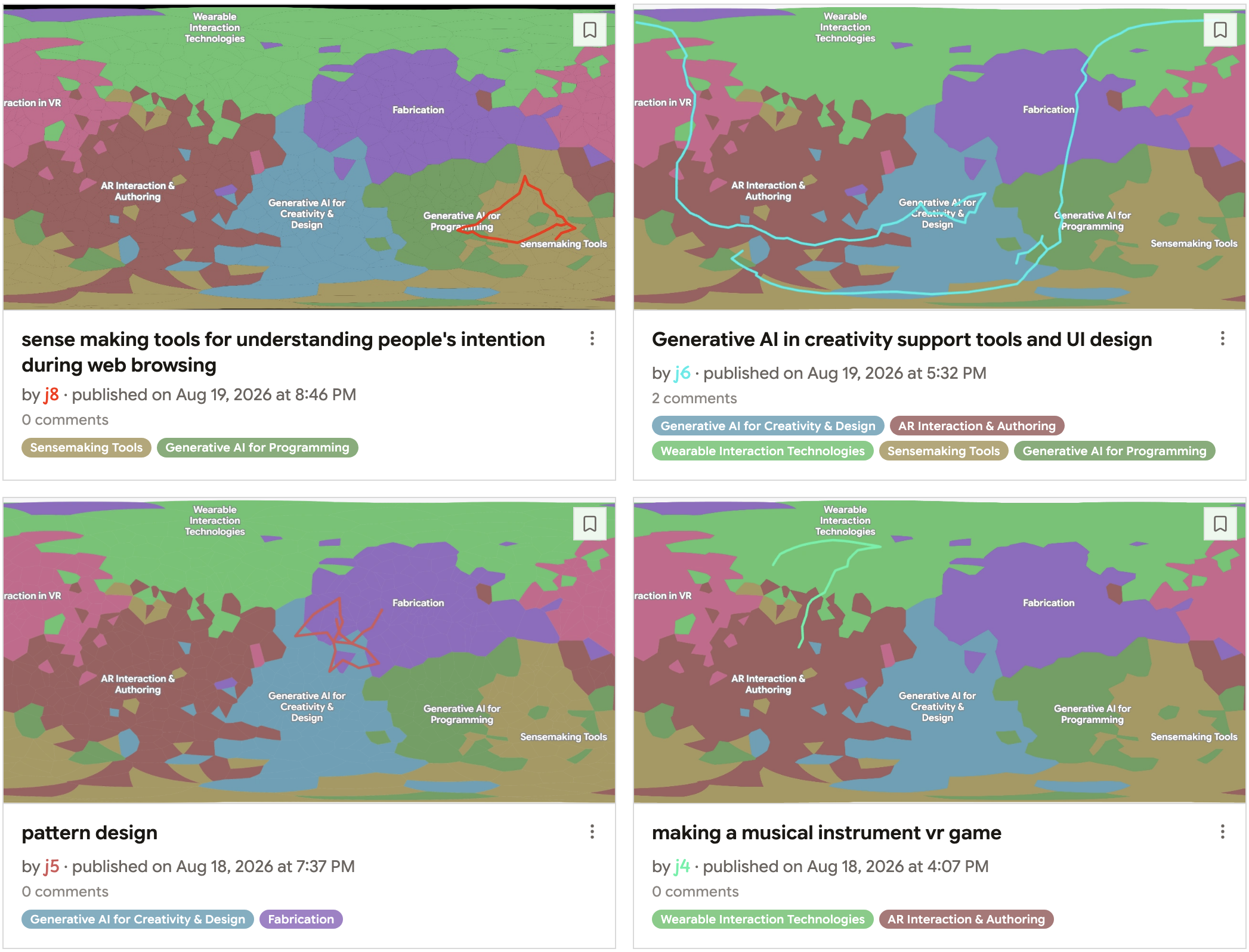}
    \caption{
Published traces in the \textit{Trace Library}, shown as cards with map thumbnails showing real users' paths through the same scholarly landscape.}
\Description{Six published reading traces displayed as cards in a three-column grid. Each card includes a map thumbnail showing the reader's trace across the shared scholarly landscape, along with its title, author, publication time, comments, and other trace metadata.}    \label{fig:tracelibrary}
\end{figure}

\begin{figure}
\centering\includegraphics[width=0.99\linewidth]{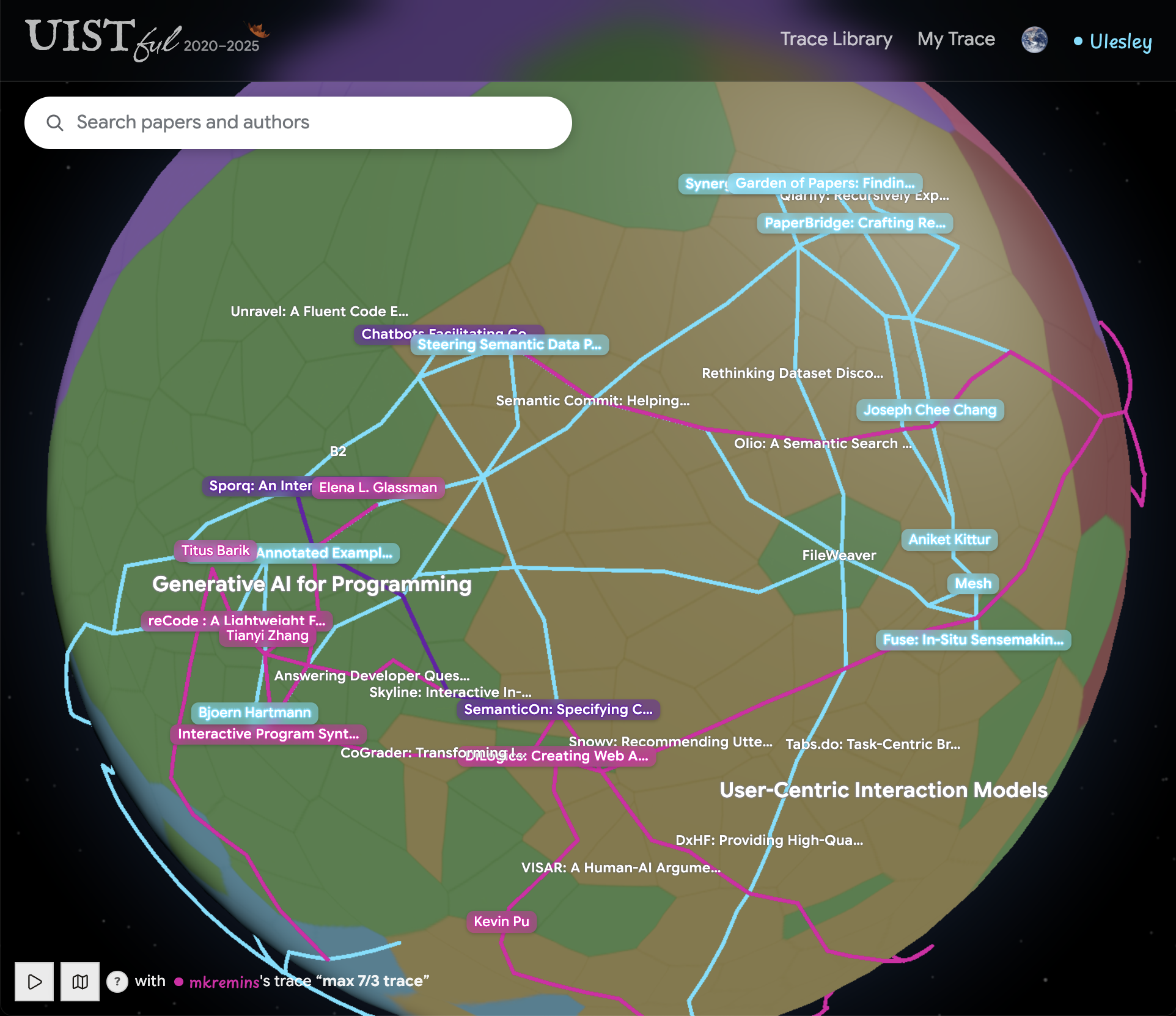}
    \caption{The Globe showing the signed-in reader's trace in light blue and a published trace in magenta. Overlapping nodes and edges appear in purple.}
    \Description{The Globe viewed against a dark background, with semantic continents shown as large colored regions and paper and author labels distributed across the surface. The signed-in reader's trace is drawn in light blue, while a published trace is drawn in magenta. Shared papers, authors, and path segments appear in purple, revealing where the two readings overlap and diverge across the globe.
}
    \label{fig:twotraces}
\end{figure}

\begin{figure*}
    \centering
    \includegraphics[width=0.98\linewidth]{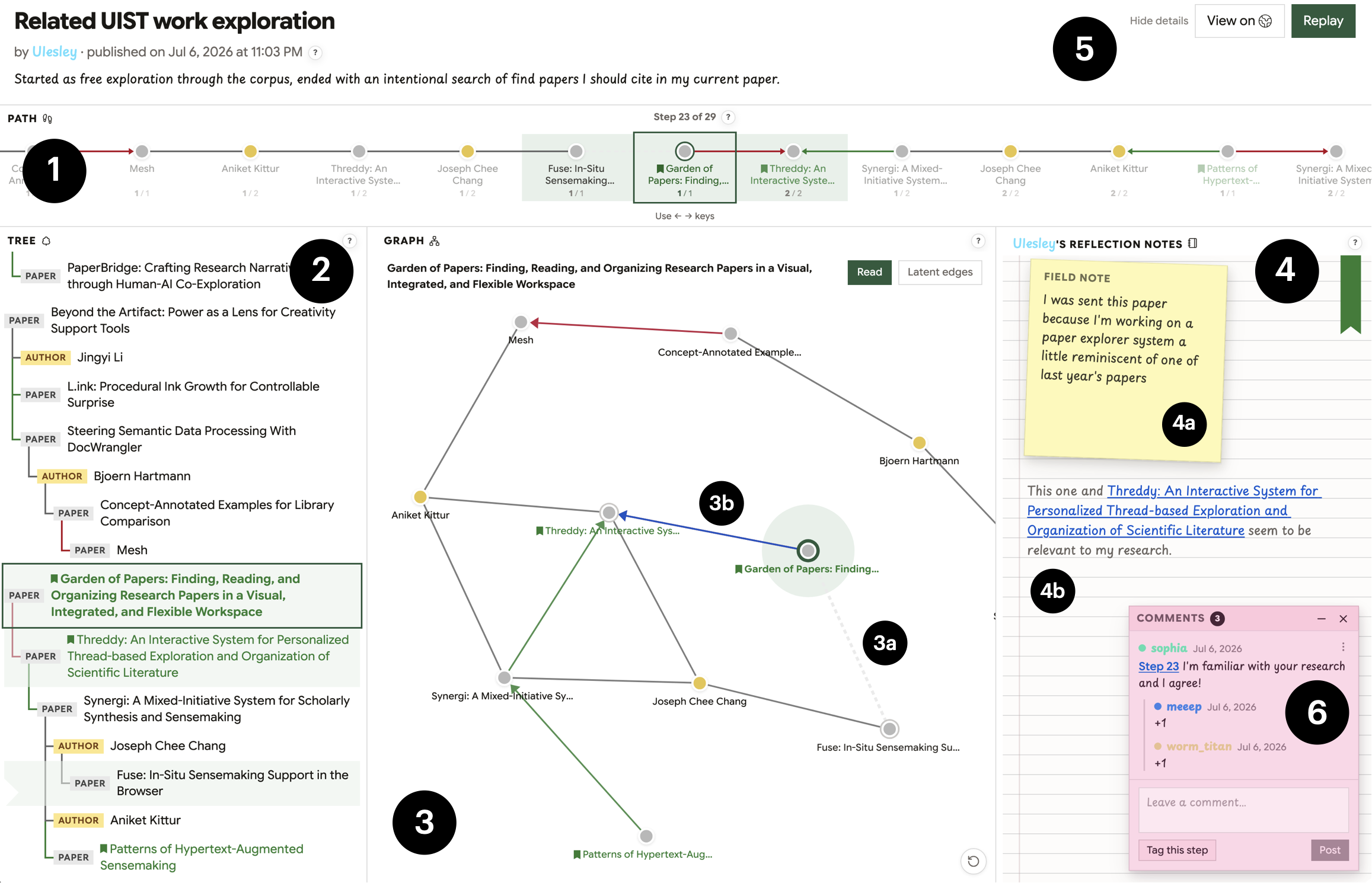}
      \caption{\textsc{UISTful}'s published trace interface presents one reading as an ordered Path (1), hierarchical Tree (2), and nonlinear Graph (3). The Graph preserves non-adjacent jumps in the trace (3a) and shows reader-authored edges (3b). In-reading \textit{Field Notes} (4a) and post-reading \textit{Reflection Notes} (4b) capture interpretation, while bookmarks highlight meaningful moments. Readers can coordinate the views, replay the trace, project it onto the Globe (5), and discuss the trace with others (6).}
  \Description{A screenshot of UISTful's published trace interface for a trace titled ``Related UIST work exploration.'' The top Path view shows a horizontal sequence of 29 visited paper and author pages, with the current item highlighted at step 23. Below, the Tree view on the left presents the trace as a hierarchy of papers and authors linked through citation and authorship relationships. The central Graph view shows the same trace as a network, preserving non-adjacent jumps in the trace and showing reader-authored connections between pages. The right Notes panel contains a yellow Field Note, a written Reflection Note, and a bookmark marker. Controls at the top allow the reader to replay the trace or view it on the Globe. A comments panel in the lower right supports discussion of the trace and individual steps.
}
\label{fig:publishedtrace}
\end{figure*}

Published traces appear in the public \textit{Trace Library}, where readers can search and filter by paper, author, user, or continent, and sort by recency or Jaccard similarity to their own trace (Fig.~\ref{fig:tracelibrary}). Opening a trace presents the Path, Tree, Graph, and notes interface from Fig.~\ref{fig:publishedtrace} in read-only form, with controls to replay it step by step through the corresponding full pages and comment on the trace or individual steps. Readers can also project a published trace onto the Globe alongside their own, making overlaps and divergences visible across the shared landscape (Fig.~\ref{fig:twotraces}).

\subsubsection{Collective Trace}
The Globe and Map offer three anonymized collective trace layers across all \textsc{UISTful} readers (Fig.~\ref{fig:maptraces}). \textit{Paths} overlays aggregate reading paths in orange, with thicker segments indicating more frequent traversal (Fig.~\ref{fig:maptraces}-b). \textit{Readers} shows each reader's route in their personal trace color, making individual paths and overlaps visible (Fig.~\ref{fig:maptraces}-c). \textit{Heat} colors paper cells by visit frequency and labels the 20 most-visited papers (Fig.~\ref{fig:maptraces}-d). By aggregating and zooming out from the individual traces in \textit{My Trace} and the \textit{Trace Library}, these layers capture a bird's-eye view of the collective creative reading happening on \textsc{UISTful}, evoking an ``Overview Effect'' at the scale of the systems research community~\cite{weWereOnline, seeingWholeEarth}.

\subsection{Usage Scenarios}

Curated traces can be saved privately or published to the community. We anticipate \textsc{UISTful} supporting both directed and exploratory reading for oneself, for others, or for both~\cite{notesToSelfToShare}.

\begin{table}[H]
    \centering

    \small
    \begin{tabular}{
        >{\raggedright\arraybackslash}p{0.18\linewidth}
        >{\raggedright\arraybackslash}p{0.35\linewidth}
        >{\raggedright\arraybackslash}p{0.35\linewidth}
    }
    \toprule
     & \textbf{For oneself} & \textbf{For others} \\
    \midrule
    \textbf{Directed} &
    Find and contextualize sources for an ongoing project.&
    Curate a path through a topic or argument. \\
    \midrule
    \textbf{Exploratory}&
    Wander for curiosity or inspiration. &
    Share an unexpected rabbithole for others to replay.\\
    \bottomrule
    \end{tabular}
    
    \vspace{0.4em}

    \caption{Potential uses of \textsc{UISTful} across reading orientation and audience.}
    \label{tab:usage-scenarios}
\end{table}


\section{Conclusion}
\begin{quote}
\textit{The photographer is an armed version of the solitary walker reconnoitering, stalking, cruising the urban inferno, the voyeuristic stroller who discovers the city as a landscape of voluptuous extremes. Adept of the joys of watching, connoisseur of empathy, the flâneur finds the world ``picturesque.''}
\end{quote}
\begin{flushright}---Susan Sontag, \textit{On Photography}\end{flushright}

We present \textsc{UISTful}, a system that turns the scholarly reader into a flâneur wandering through a shared landscape of papers and authors. Reading traces serve as a camera through which readers frame and interpret what they encounter, much as a flâneur interprets a city through photography, demonstrating reading as a creative process~\cite{painterOfModern, onPhotography, creativeReading}. Shared traces expose a plurality of readings, while collective views induce an overview effect through which UIST, a conference on interactive systems, can see itself as an interactive system of papers, authors, readers, and their exchanges. As participants read, reflect, and share, they compose a living portrait of their scholarly community, enacting the participant--observer and reader--author dialectics at the heart of \textsc{UISTful}.



\begin{acks}
We thank Kosa Goucher-Lambert, Hugh Dubberly, TJ McLeish, John Cain, Kevin Ma, and Caseysimone Ballestas of the knowledge graph discussion group, as well as Ronen Tamari and Wesley Finck of Semble and the Cosmik Network, for thoughtful discussions, feedback, and inspiration for the system.
\end{acks}

\bibliographystyle{ACM-Reference-Format}
\bibliography{bibliography}

@String{Computing = "Computing" }

@String{Macmillan = "Macmillan" }

@inproceedings{gardenOfPapers,
author = {Ma, Donghyeok and Jang, Hanbee and Lee, Joon Hyub and Bae, Seok-Hyung},
title = {Garden of Papers: Finding, Reading, and Organizing Research Papers in a Visual, Integrated, and Flexible Workspace},
year = {2025},
isbn = {9798400720376},
publisher = {Association for Computing Machinery},
address = {New York, NY, USA},
url = {https://doi.org/10.1145/3746059.3747637},
doi = {10.1145/3746059.3747637},
booktitle = {Proceedings of the 38th Annual ACM Symposium on User Interface Software and Technology},
articleno = {89},
numpages = {15},
location = {
},
series = {UIST '25}
}

@misc{weWereOnline, url={https://wewere.online/}, journal={we were online}, author={Chang, Spencer}, year={2026}}

@article{SemanticReader,
author = {Lo, Kyle and Chang, Joseph Chee and Head, Andrew and Bragg, Jonathan and Zhang, Amy X. and Trier, Cassidy and Anastasiades, Chloe and August, Tal and Authur, Russell and Bragg, Danielle and Bransom, Erin and Cachola, Isabel and Candra, Stefan and Chandrasekhar, Yoganand and Chen, Yen-Sung and Cheng, Evie Yu-Yen and Chou, Yvonne and Downey, Doug and Evans, Rob and Fok, Raymond and Hu, Fangzhou and Huff, Regan and Kang, Dongyeop and Kim, Tae Soo and Kinney, Rodney and Kittur, Aniket and Kang, Hyeonsu B. and Klevak, Egor and Kuehl, Bailey and Langan, Michael J. and Latzke, Matt and Lochner, Jaron and MacMillan, Kelsey and Marsh, Eric and Murray, Tyler and Naik, Aakanksha and Nguyen, Ngoc-Uyen and Palani, Srishti and Park, Soya and Paulic, Caroline and Rachatasumrit, Napol and Rao, Smita and Sayre, Paul and Shen, Zejiang and Siangliulue, Pao and Soldaini, Luca and Tran, Huy and van Zuylen, Madeleine and Wang, Lucy Lu and Wilhelm, Christopher and Wu, Caroline and Yang, Jiangjiang and Zamarron, Angele and Hearst, Marti A. and Weld, Daniel S.},
title = {The Semantic Reader Project},
year = {2024},
issue_date = {October 2024},
publisher = {Association for Computing Machinery},
address = {New York, NY, USA},
volume = {67},
number = {10},
issn = {0001-0782},
url = {https://doi.org/10.1145/3659096},
doi = {10.1145/3659096},
journal = {Commun. ACM},
month = sep,
pages = {50–61},
numpages = {12}
}

@online{Semble,
  author    = {{Semble}},
  title     = {Semble: A Social Knowledge Network for Creators},
  year      = {2025},
  url       = {https://semble.so},
  urldate   = {2026-02-20}
}

@misc{creativeReading,
      title={Creative Reading: Scaffolding Reading for Transformation}, 
      author={Sophia Liu and Sarah Abowitz and Yijun Liu and Sarah Sterman and Shm Garanganao Almeda and Max Kreminski},
      year={2026},
      eprint={2606.04308},
      archivePrefix={arXiv},
      primaryClass={cs.HC},
      url={https://arxiv.org/abs/2606.04308}, 
}

@inproceedings{goingDownTheRabbithole,
author = {Piccardi, Tiziano and Gerlach, Martin and West, Robert},
title = {Going Down the Rabbit Hole: Characterizing the Long Tail of Wikipedia Reading Sessions},
year = {2022},
isbn = {9781450391306},
publisher = {Association for Computing Machinery},
address = {New York, NY, USA},
url = {https://doi.org/10.1145/3487553.3524930},
doi = {10.1145/3487553.3524930},
booktitle = {Companion Proceedings of the Web Conference 2022},
pages = {1324–1330},
numpages = {7},
location = {Virtual Event, Lyon, France},
series = {WWW '22}
}

@inproceedings{informationCity,
author = {Bernstein, Mark and Hooper, Silas and Anderson, Mark W. R.},
title = {Back To The Information City},
year = {2025},
isbn = {9798400715341},
publisher = {Association for Computing Machinery},
address = {New York, NY, USA},
url = {https://doi.org/10.1145/3720553.3746664},
doi = {10.1145/3720553.3746664},
booktitle = {Proceedings of the 36th ACM Conference on Hypertext and Social Media},
pages = {118–126},
numpages = {9},
location = {
},
series = {HT '25}
}

@inproceedings{informationFlaneur,
author = {D\"{o}rk, Marian and Carpendale, Sheelagh and Williamson, Carey},
title = {The information flaneur: a fresh look at information seeking},
year = {2011},
isbn = {9781450302289},
publisher = {Association for Computing Machinery},
address = {New York, NY, USA},
url = {https://doi.org/10.1145/1978942.1979124},
doi = {10.1145/1978942.1979124},
booktitle = {Proceedings of the SIGCHI Conference on Human Factors in Computing Systems},
pages = {1215–1224},
numpages = {10},
location = {Vancouver, BC, Canada},
series = {CHI '11}
}

@inproceedings{Sensecape,
author = {Suh, Sangho and Min, Bryan and Palani, Srishti and Xia, Haijun},
title = {Sensecape: Enabling Multilevel Exploration and Sensemaking with Large Language Models},
year = {2023},
isbn = {9798400701320},
publisher = {Association for Computing Machinery},
address = {New York, NY, USA},
url = {https://doi.org/10.1145/3586183.3606756},
doi = {10.1145/3586183.3606756},
booktitle = {Proceedings of the 36th Annual ACM Symposium on User Interface Software and Technology},
articleno = {1},
numpages = {18},
location = {San Francisco, CA, USA},
series = {UIST '23}
}

@inproceedings{Threddy,
author = {Kang, Hyeonsu and Chang, Joseph Chee and Kim, Yongsung and Kittur, Aniket},
title = {Threddy: An Interactive System for Personalized Thread-based Exploration and Organization of Scientific Literature},
year = {2022},
isbn = {9781450393201},
publisher = {Association for Computing Machinery},
address = {New York, NY, USA},
url = {https://doi.org/10.1145/3526113.3545660},
doi = {10.1145/3526113.3545660},
booktitle = {Proceedings of the 35th Annual ACM Symposium on User Interface Software and Technology},
articleno = {94},
numpages = {15},
location = {Bend, OR, USA},
series = {UIST '22}
}

@inproceedings{Artographer,
author = {Almeda, Shm Garanganao and Chung, John Joon Young and Liu, Sophia and Lu, Yuwen and Halperin, Brett and Hartmann, Bjoern and Kreminski, Max},
title = {Artographer: a Curatorial Interface for Art Space Exploration},
year = {2026},
isbn = {9798400725838},
publisher = {Association for Computing Machinery},
address = {New York, NY, USA},
url = {https://doi.org/10.1145/3803784.3807532},
doi = {10.1145/3803784.3807532},
booktitle = {Proceedings of the 2026 Conference on Creativity and Cognition},
pages = {377–396},
numpages = {20},
location = {
},
series = {C\&C '26}
}

@online{seeingWholeEarth,
  author    = {{Ahmed Kabil}},
  title     = {Seeing the Whole Earth From Space Changed Everything},
  url       = {https://longnow.org/ideas/whole-earth-overview-effect/},
  date   = {2018-05-29}
}

@inproceedings{footprintsMaes,
author = {Wexelblat, Alan and Maes, Pattie},
title = {Footprints: history-rich tools for information foraging},
year = {1999},
isbn = {0201485591},
publisher = {Association for Computing Machinery},
address = {New York, NY, USA},
url = {https://doi.org/10.1145/302979.303060},
doi = {10.1145/302979.303060},
booktitle = {Proceedings of the SIGCHI Conference on Human Factors in Computing Systems},
pages = {270–277},
numpages = {8},
location = {Pittsburgh, Pennsylvania, USA},
series = {CHI '99}
}

@inproceedings{CATsDesignSpace,
author = {Hammad, Noor and Lin, David Chuan-En and Smith, Amy and Kreminski, Max and Harpstead, Erik and Hammer, Jessica},
title = {Tracing Creativity: A Design Space For Creative Activity Traces in HCI},
year = {2026},
isbn = {9798400722783},
publisher = {Association for Computing Machinery},
address = {New York, NY, USA},
url = {https://doi.org/10.1145/3772318.3791263},
doi = {10.1145/3772318.3791263},
booktitle = {Proceedings of the 2026 CHI Conference on Human Factors in Computing Systems},
articleno = {1644},
numpages = {24},
location = {
},
series = {CHI '26}
}

@inproceedings{notesToSelfToShare,
author = {Crescenzi, Anita and Li, Yuan and Zhang, Yinglong and Capra, Rob},
title = {Towards Better Support for Exploratory Search through an Investigation of Notes-to-self and Notes-to-share},
year = {2019},
isbn = {9781450361729},
publisher = {Association for Computing Machinery},
address = {New York, NY, USA},
url = {https://doi.org/10.1145/3331184.3331309},
doi = {10.1145/3331184.3331309},
booktitle = {Proceedings of the 42nd International ACM SIGIR Conference on Research and Development in Information Retrieval},
pages = {1093–1096},
numpages = {4},
location = {Paris, France},
series = {SIGIR'19}
}

@book{onPhotography,
  title     = {On Photography},
  author    = {Sontag, Susan},
  year      = {1977},
  publisher = {Farrar, Straus and Giroux},
  address   = {New York},
  isbn      = {978-0-312-42009-3}
}

@inproceedings{abstractExplorer,
author = {Gu, Ziwei and Zhou, Joyce and Lei, Ning-Er (Nina) and Kummerfeld, Jonathan K. and Jasim, Mahmood and Mahyar, Narges and Glassman, Elena L.},
title = {AbstractExplorer: Leveraging Structure-Mapping Theory to Enhance Comparative Close Reading at Scale},
year = {2025},
isbn = {9798400720376},
publisher = {Association for Computing Machinery},
address = {New York, NY, USA},
url = {https://doi.org/10.1145/3746059.3747773},
doi = {10.1145/3746059.3747773},
booktitle = {Proceedings of the 38th Annual ACM Symposium on User Interface Software and Technology},
articleno = {85},
numpages = {25},
location = {
},
series = {UIST '25}
}

@book{painterOfModern,
  title     = {The Painter of Modern Life and Other Essays},
  author    = {Charles Baudelaire},
  editor    = {Jonathan Mayne},
  year      = {1964},
  publisher = {Phaidon Press},
  address   = {London},
  isbn      = {0714812668}
}

@online{GoogleEarth,
  author    = {{Google}},
  title     = {Google Earth},
  year      = {2026},
  url       = {https://earth.google.com/},
  urldate   = {2026-07-08}
}

\end{document}